\newcommand{\br}{{\bf r}}
\newcommand{\bd}{{\bf d}}

\newcommand{\be}{{\bf e}_{{\bf r}}}

\documentstyle[aps,twocolumn]{revtex}
\begin{document}

\wideabs{

\title{On the applicability of the classical dipole-dipole
 interaction for polar Bose-Einstein condensates.}

\author{Vladimir V. Konotop \cite{email1}}

\address{Departamento de F\'{\i}sica e
Centro de F\'{\i}sica da Mat\'eria Condensada,  Universidade de Lisboa,
Complexo Interdisciplinar, \\ Av.
Prof. Gama Pinto, 2, Lisbon, P-1649-003 Portugal}

\author{V\'{\i}ctor M. P\'erez-Garc\'{\i}a \cite{email2}}

\address{Departamento de Matem\'aticas, E.T.S.I. Industriales, Universidad de Castilla-La Mancha,
\\ Avda. Camilo Jos\'e Cela, 3, Ciudad Real, 13071 Spain.}

\date{\today}
\maketitle
\begin{abstract}
We argue that the classical form of the dipole-dipole interaction
energy cannot be used to model the interaction of the bosons in a
dilute Bose-Einstein condensate made of polar atoms. This fact is
due to convergence of integrals, if no additional restrictions are
introduced. The problem can be regularized, in particular, by
introducing a hard sphere model. As an example we propose a
regularization consistent with the long range behavior of the
effective potential and with the scattering amplitude of the fast
particles.

\end{abstract}

}

\narrowtext


The first clear observation of pure Bose-Einstein condensation was done using
a ultracold  gas of neutral bosonic atoms
\cite{siempre}. After these remarkable experiments an explosive growth of
the interest on the subject
happened \cite{web}. Following the experimental guidelines, the theoretical
 analysis of the problem concentrated on
 condensates made of alkaline atoms (neutral and nonpolar) interacting collisionally
through $s$-wave spherically symmetric effective potentials, which
lead to an
 interaction proportional to the local
atomic density.

Although this description of the problem captures most of what has
been done experimentally up to now, very recently there has been
increased interest on other, more complex, interactions.
Specifically, there has been some interest on nonlocal
interactions  (see e.g. \cite{nuestro} and references therein). In
fact any realistic interaction should always be nonlocal  due to
the fact that the range of interaction is not zero. Another
situation corresponds to nonsymmetric and nonlocal interactions
considered in Refs. \cite{You,Goral,Dell,Santos}, such as those
which appear when the bosons inside the condensate are polar
molecules (with permanent or externally induced dipole moments).
The possibility of finding condensates with nonsymmetric
interactions is very interesting since it would lead to a bunch of
new phenomena and many possibilities for control.

In all the previously cited cases the model for the dynamics of
the system is the zero temperature mean field theory, i. e.
a Gross-Pitaevskii equation, for the condensate dynamics

\begin{eqnarray}
\label{schroe} i\hbar \frac{\partial \Psi}{\partial t} = -
\frac{\hbar^2}{2m} \triangle \Psi + \frac{1}{2}m\omega_0^2\left(
x^2 + y^2 + \gamma^2 z^2 \right) \Psi \nonumber\\
+ \frac{4\pi\hbar^2a}{m} \left|\Psi\right|^2 \Psi +
\left(\int V({\bf r},{\bf r}') \left|\Psi({\bf r}')\right|^2
d{\bf r}' \right) \Psi.
\end{eqnarray}
where $\Psi$ is the condensate wavefunction normalized to the
total number of particles,
\begin{equation}
N = \int d{\bf r} \left|\Psi({\bf r})\right|^2,
\end{equation}
$a$ is the s-wave scattering length and $V$ is a real function
taking into account the nonlocal part of the interactions. Of
course, $V$ vanishes or is very small when the interactions are
purely local. When the interactions are mediated by dipoles, they
are supposed to be long range.  In such situation it is proposed
that the interaction is ruled by a potential of the type
\cite{You,Goral,Santos}
 \begin{equation}
\label{dipole-dipole} V({\bf r},{\bf r'})= \frac{\mu_0}{4\pi}
\frac{\bd({\bf r}) \bd({\bf r'})-3(\bd({\bf r}) \be)(\bd({\bf
r'})\be )}{|{\bf r}-{\bf r'}|^3}
\end{equation}
where $\be =({\bf r}-{\bf r'})/|{\bf r}-{\bf r'}|$ and
$\bd({\bf r})$ is the dipole moment at point ${\bf r}$.   The simplest
version of Eq. (\ref{dipole-dipole}), where $\bd({\bf r})\equiv
{\bf d}$ is homogeneous and the potential can be considered as a function of relative distances only, has the form
\begin{equation}
\label{dipole-dipole2} V({\bf r}-{\bf r'})= \frac{\mu_0d^2}{4\pi}
\frac{1-3\cos^2 \theta}{|{\bf r}-{\bf r'}|^3},
\end{equation}
where $\cos \theta = {\bf e}_{\bd}\cdot \be$.  In some cases
\cite{You}, instead of the simple angular dependence $1-3\cos^2
\theta$, which is proportional to the spherical harmonic $Y_{20}$
more complex dependencies in the form of a finite combination of
spherical harmonics may appear.

It is our intention in this brief report to show that all those
potentials lead to a singular (in mathematical sense) interaction
which must be regularized to give physically acceptable results.
This situation is not surprising if one takes into account that
even in classical theory (i) Eq. (\ref{dipole-dipole}) is an
asymptotics of the dipole-dipole interactions (see e.g. \cite{LL})
and thus fails in the near zone, and (ii) Eq.
(\ref{dipole-dipole}) displays the realistic dipole-dipole
interaction while $V({\bf r})$ in (\ref{schroe}) is an effective
potential obtained within the mean field theory.



To simplify the analysis, let us first consider the solution of
the Gross-Pitaevskii equation with an interaction term of the form
(\ref{dipole-dipole2}). Then the nonlinear nonlocal term is
proportional to the integral
\begin{equation}
\label{int} I({\bf r})  = \int d{\bf r'} \frac{1-3\cos^2\theta}{|{\bf
r}-{\bf r'}|^3} J({\bf r'})
\end{equation}
being $J({\bf r'}) = |\Psi({\bf r'})|^2$.  Taking into account
that due to the presence of the trap and the finite norm of the
solution the wave function decays as $|{\bf r}|\to \infty$, the
only point where the integral may lead to a singular behavior is
${\bf r'}={\bf r}$, thus we will employ spherical coordinates and
estimate the integral in the $\epsilon-$vicinity of ${\bf u}\equiv
{\bf r}-{\bf r'}$. To do so we split (hereafter
$d\Omega=\sin\theta d\theta d\phi$)
\begin{eqnarray}
I({\bf r}) & = &  \left(\int_0^{\epsilon} + \int_{\epsilon}^{\infty} \right) \frac{du}{u} \int d\Omega
(1-3\cos^2\theta) \, J({\bf u}+{\bf
r}) \nonumber \\ & =& I_{\epsilon}({\bf r}) + Q({\bf r}).
\end{eqnarray}
The singular behavior, if present, may only be due to the
contribution of $I_{\epsilon}({\bf r})$ for which we have the
following set of estimates
\begin{eqnarray}
\label{expan} I_{\epsilon}  &=&   \int_0^{\epsilon} \frac{du}{u} \int d\Omega
(1-3\cos^2\theta) \, J({\bf u}+{\bf
r}) \nonumber \\
&=& \int_0^{\epsilon} \frac{du}{u} \int d\Omega(1-3\cos^2\theta) \,
\left[J({\bf r}) + \left({\bf u}\nabla \right) J({\bf r}) + {\cal
O}(u^2)\right]
\nonumber \\
&=& \int_0^{\epsilon} \frac{du}{u} \int d\Omega (1-3\cos^2\theta)
\,J({\bf r})+{\cal O}(1).
\end{eqnarray}
Taking into account that $J({\bf r})$ does not depend on the
integration variable one gets an indetermination of type $0
\times \infty$ which cannot be avoided because the order of
integration is not defined (and should not be relevant in a
physically meaningful model). This is the first indication that
there may be problems with the type of nonlocal interaction
kernels considered.

To rule out the possibility that the singular behavior previously
found is a product of the specific orientation of the dipoles
chosen (i.e. all of them
 parallel) we consider now a general situation. Let us first define a new
function
\begin{equation} {\bf \Delta} ({\bf r},{\bf u}) = {\bf
d}({\bf r}+{\bf u})-{\bf d}({\bf r}). \end{equation}
 With this
definition we may write
\begin{eqnarray}
I({\bf r}) & = & \int d{\bf u} J\left({\bf r}+{\bf u}\right) \nonumber \\
& \times & \frac{\bd({\bf r}) \bd({\bf r} + {\bf u})-3(\bd({\bf r}) \be )(\bd({\bf
r}+{\bf u})\be)}{u^3} \nonumber \\
& = & \int \frac{d{\bf u}}{u^3} J\left({\bf r}+{\bf u}\right)
\left[\bd({\bf r})\bd({\bf r})-3(\bd({\bf r})\be)(\bd({\bf r})\be) \right. \nonumber \\
& & +   \left. {\bf \Delta}({\bf r},{\bf u}) {\bf d}({\bf r}) -
3({\bf \Delta}({\bf r},{\bf u}) \be)({\bf d}({\bf r})
\be) \right].
\end{eqnarray}
Let us formally split this integral into two parts, $I({\bf r}) = I_0({\bf r}) + I_1({\bf r})$. The first one
\begin{eqnarray}
I_0({\bf r}) & =  & \int \frac{d{\bf u}}{u^3} J\left({\bf r}+{\bf u}\right)
\left[\bd({\bf r})\bd({\bf r})-3(\bd({\bf r})\be)(\bd({\bf r})\be) \right] \nonumber \\
& =& d^2({\bf r}) \int \frac{d{ u}}{u} d\Omega J\left({\bf r}+{\bf u}\right) \left[1-3\cos^2 \theta \right]
\end{eqnarray}
is of the same type as Eq. (\ref{int})) and leads to an
indetermination. The second contribution to $I({\bf r})$ is
\begin{equation}
\label{covn}
I_1({\bf r})  = {\bf d}({\bf r}) \int \frac{du}{u}d\Omega J({\bf r}+{\bf u})\left[
 {\bf \Delta}({\bf r},{\bf u})  - 3({\bf \Delta}({\bf r},{\bf u}) \be)  \be\right].
\end{equation}
Let us define
\begin{equation}
\label{cov2}
I_2({\bf r})  =  \int \frac{du}{u}d\Omega J\left({\bf r}+{\bf u}\right) {\bf \Delta}({\bf r},{\bf u}).
\end{equation}
Then we may write
\begin{equation}
\label{covn2}
I_1({\bf r})  = {\bf d}({\bf r}) \left[ I_2({\bf r}) - 3(I_2({\bf r}) \be)  \be\right].
\end{equation}
Thus if (\ref{cov2}) is convergent one may ensure convergence of (\ref{covn}) or equivalently (\ref{covn2}).
Assuming that ${\bf d}({\bf r})$ is a differentiable function and expanding in
Taylor series we find that ${\bf \Delta}({\bf r},{\bf u}) = {\cal{O}}(u)$ and
thus the singular behavior for small $u$ values is avoided. Thus the convergence
of the integral depends on the contribution of $I_0({\bf r})$. This means that the
consideration in what follows can be restricted to the model (\ref{dipole-dipole2})
 without restriction of generality.


As we have commented above the integral $I_0({\bf r})$ is not well
defined. Let us show, however that the problem may be regularized.
Indeed, assuming that the dipole-dipole interaction is cut off at
some distance $a_c$ one has to substitute the lower limit in
(\ref{expan}) by $a_c$ what will lead to the result of integration
which is not divergent at $a_c\to 0$. Mathematically
\begin{equation}
\label{e31} I_0({\bf r}) = d^2({\bf r}) \int_{a_c}^{\infty}
\frac{d{u}}{u} \int d\Omega
 J\left({\bf r}+{\bf u}\right) \left[1-3\cos^2 \theta \right]
\end{equation}
is uniformily bounded with respect to ${\bf r}$. The contribution
of the small scale of the integral can be estimated
 as before by expanding $J({\bf r})$ as
\begin{eqnarray}
\int_{a_c}^{\epsilon} \frac{du}{u} \int d\Omega (1-3\cos^2\theta)
\,J({\bf r})+{\cal O}(1) \nonumber \\
 = J({\bf r}) \left(\log a_c \right) \int d\Omega (1-3\cos^2\theta)+{\cal O}(1) = {\cal O}(1),
\end{eqnarray}
In fact one may easily prove that the result may be expanded in power series of $a$ as
\begin{equation}
I_0({\bf r}) = I_0^0({\bf r}) + a_c I_0^1({\bf r}) + ...
\end{equation}
so that, in the limit $a_c\rightarrow 0$ one gets a finite value for $I_0({\bf r})$.

The fact that the integral is well behaved when a cutoff is used is the reason
 why no divergences were observed in numerical studies \cite{You,Goral,Dell}.
The use of any computational mesh introduces a cutoff related to the mesh size
which is effectively equivalent to introducing the hard sphere radius.

The cut-off model, although the most straightforward way of
avoiding singularities of the problem is not the most natural one,
from the physical point of view. Indeed, this is not properly a
``hard sphere" morel which is often used in physics, since the
  respective cut-off is not introduced in the self-interaction term.
  Thus such a cut-off corresponds to zero dipole-dipole interaction
   at small distances which is not correct
   from the physical point of view.

In order to propose another way of regularization in the case at hand we take
into account that in the field theory dipoles appear as an artifact of the energy
 expansion with respect to a small parameter $r_d/r$ where $r_d$ is a characteristic
 dimension of the dipole (for the next terms of the expansion one has to compute
multipolar terms). The complete potential energy is just made of two-particle Coulomb
interactions. This imposes the  first requirement: the singularity of the potential
must be of the $1/r$-type when the distance between two particles goes to zero.
The second requirement to the model naturally follows from
 the fact that although one cannot compute the potential exactly, it must have at
least one ``free" parameter which describes the size of the transition region between
 $1/r$ and $1/r^3$ laws (this is the parameter which substitutes the scattering
length arising in the theory including only local interactions).
The third requirement is that in the limit $r\to\infty$ the
potential must acquire the form (\ref{dipole-dipole2}). Finally,
the potential must be differentiable in the whole space (except
the origin). A simple form
 which satisfies all these requirements is
\begin{equation}
\label{dipole-dipole3} V({\bf r})= -\frac{\mu_0d^2}{4\pi}
 \left(\exp\left(-\frac{r^2}{a_d^2}\right)-1\right)  \frac{1-3\cos^2 \theta}{r^3}
\end{equation}
where $a_d$ is a constant.


A natural question which appears after the regularized potential is
introduced is how it affects the results obtained, so far. First of all
 we mention that (\ref{schroe}) with the potential (\ref{dipole-dipole2})
 without cutoff does not allow plane wave solutions, just because the
respective integrals diverge (and not because of instability). Introducing
a cutoff does not change the situation since it makes the equation explicitly spatially dependent
and so it still has no plane wave solutions.

 It is easy to see that one of the main effects of regularization is the effect on
 the collapse or blow-up phenomenon.
Namely, we are going to  show now that any kind of cutoff prevents collapse. First however,
 we point out a heuristic argument in favor of our statement. The model (\ref{dipole-dipole}),
 which is not regularized, is formally invariant with respect to  the renormalization
$\Psi\mapsto \Psi/L$, $\br \mapsto L\br$, $t\mapsto L^2 t$ (see e.g. \cite{Sulem})
while any kind of regularization breaks this symmetry indicating the possibility
of existence of a ground state solution. Such a solution (if any) will be stable if the
respective Hamiltonian has a lower bound at constant number of particles.

So we now proceed to prove that such a bound in the case of
positive (or zero) scattering length exists \cite{com4}. As a
matter of fact after introducing a regularized potential the proof
of the stability is reduced to one given in \cite{Turitsyn}.
Indeed considering Eq. (\ref{schroe}) and using appropriate
dimensionless variables one can write down an associated
Hamiltonian in the form
\begin{eqnarray}
\label{hamilt} H&=&\int \left(|\nabla \psi|^2+|\psi|^4 + V({\bf
r}) |\psi|^2\right) d\br
\nonumber \\
&-&\int d\br\int d\br'  |\psi(\br)|^2 |\psi(\br')|^2\frac{f(\br-\br')}{|\br-\br'|}
\end{eqnarray}
Any regularization satisfying the conditions imposed above
satisfies also the constraint
\begin{equation}
\label{req}
|f(\br)|\leq f_0<\infty,
\end{equation}
where $f_0=$const. For instance this condition is strictly
satisfied by  regularization (\ref{dipole-dipole3}). We would like
to stress the essential fact that the particular function $f(\br)$
used to regularize is not essential provided Eq. (\ref{req}) is
satisfied (for example, the cut-off model also satisfies this
requirement, as well). Next we will use the fact that $\int V({\bf
r}) |\psi|^2 d\br >0$ and the inequality \cite{Turitsyn}
\begin{eqnarray}
\label{estim2}
\int d\br\int d\br'  |\psi(\br)|^2 |\psi(\br')|^2\frac{f(\br-\br')}{|\br-\br'|}
\nonumber \\
\leq 2 f_0 N^{3/2}\left(\int|\nabla\psi|^2d\br\right)^{1/2}
\end{eqnarray}
Finally
\begin{equation}
H \geq \|\nabla \psi\|_2^2-2f_0 \|\nabla \psi\|_2\|
 \psi\|_2^3  \geq -f_0^2N^3,
\end{equation}
and thus the hamiltonian is bounded below, i.e.  strict collapse
is not possible. Of course some tendency of the system to compress
would be observed in the dynamics.


 To conclude, in this report we have shown that the classical dipole-dipole
  interaction given by Eq. (\ref{dipole-dipole}) is not consistent and must
  be regularized somehow to take into account the divergence near the origin.
  Fortunately the model is well behaved in the sense that it is possible to
  regularize in several reasonable (and physically meaningful) ways. In fact
  the usual numerical treatment of the problem includes implicitly a regularization
   which is why previous works with this interactions did not show up the
   bad-possedness of the nonregularized model.

 Finally we have shown that any reasonable choice of the regularization
 leads to a supression of the collapse in the sense that the ground state
 of Eq. (\ref{schroe}) exists. Of course the effect of other physical terms
 should be taken into consideration and depending on the scale at which collapse
  is stopped the tendency to srink the solutions could lead to an effective
  depletion of the condensate even in this case in which strict collapse is not possible.

\acknowledgements

Authors are grateful to G. Alfimov for useful comments. VVK
acknowledges support from FEDER and Program PRAXIS XXI, No
Praxis/P/Fis/10279/1998. V.M.P-G. has been supported by Ministerio
de Ciencia y Tecnolog\'{\i}a
under grant BFM2000-0521. The cooperative work has been supported
through the bilateral program DGCYT-HP1999-019/A\c{c}\~{a}o No.
E-89/00.


\begin{references}
\bibitem{com4} This results means that the instability and collapse reported
in \cite{Goral}  and \cite{Santos} are related to the anomalously
strong singularity (of $1/r^3$ type) of the potential at the
origin. As we have commented before this behavior is nonphysical
for small $r$ and collapse would probably be stopped. This fact
does not rule out however  the possible existence of some
shrinking tendency of the atomic cloud for some spatial scales.
\end{references}
\end{document}